\documentclass[12pt]{article}
\usepackage{hyperref}

\usepackage{graphicx}
\graphicspath{ {./images/} }

\newenvironment{sciabstract}{%
\begin{quote} }
{\end{quote}}

\title{Lessons From  Model Risk Management in Financial Institutions for Academic Research} 

\author{Mahmood Alaghmandan,$^{1,}$\footnote{Corresponding Author.}  \hspace{1mm}  Olga Streltchenko$^{2}$\\
\\
\normalsize{\it $^{1}$Specialist, Climate Scenario Analysis, Climate Risk Hub}\\
\normalsize{\it $^{2}$Director, Climate Scenario Analysis, Climate Risk Hub}\\
\normalsize{Office of the Superintendent of Financial Institutions,}\\
\normalsize{ 255 Albert Street,Ottawa, Ontario K1A 0H2 Canada}\\
\small{$^{1}${\bf Email:} \texttt{mahmood.alaghmandan@osfi-bsif.gc.ca},}\\
\small{$^{2}${\bf Email:} \texttt{olga.streltchenko@osfi-bsif.gc.ca} }
}

\date{\today}

\begin{document} 


\baselineskip24pt


\maketitle

{
\setlength\parindent{150pt} 
{\it  ``You wouldn't like it here.

There is no entertainment,

and the critics are severe."}

\setlength\parindent{190pt} 
{\tiny \ \ \ \ \ Leonard Cohen (1934-2016)}
}


\begin{sciabstract}
  {\bf Abstract:} {In this paper, we discuss aspects of model risk management in financial institutions  which could be adopted by academic institutions  to  improve the  process of conducting academic research, identify and mitigate existing limitations, decrease the possibility of erroneous results, and prevent fraudulent activities. }
\end{sciabstract}


\section{Introduction}

In light of recent allegations of misconduct concerning highly regarded academic research projects in behavioural science (see \cite{plnetMoney} and \cite{newYorker}) we review model risk management practices commonly applied in financial institutions. We maintain these are transferable to academic research with clear benefits for the quality of the research outcomes. We illustrate how model governance could improve the quality of academic research, identify and mitigate limitations, reduce the occurrence of erroneous results, and prevent fraudulent activities in academia. 

By a \emph{model}, we mean a quantitative representation of the essential structure of some relationship, object or event used to emulate current reality and/or project potential future outcomes. Modelling is rooted in real world data (see Section~2 in \cite{bennett}) and thus is an inductive undertaking. Here we exclude deductive academic fields, such as abstract mathematics or theoretical physics, from consideration. While increased scrutiny that comes with structured model governance benefits all academic efforts, it is specifically the data-driven research that is susceptible to bias and error. In the following,  `academic research' refers to inductive  projects, unless otherwise is stated. 

In finance, models are being used for decision support in almost every line of business, from determining credit worthiness of a potential obligor to flagging  transactions suspected for money laundering. A financial institution uses a wide range of models to function properly.  While the outputs of these suites of models support financial decisions, the constituent models themselves produce economic, behavioural, climate, and financial estimates. Since models capture a simplified version of reality, there is risk associated with their use due to their limitations. As once said by a British statistician, George Box (1919-2013), ``all models are wrong, some are useful.” In general, \emph{model risk} is defined as ''the risk of loss arising from decisions based on incorrect or misused model outputs'' \cite{feds}.

Any model error would make a financial institution susceptible to financial, legal, and reputational losses. To mitigate these risks, financial institutions manage them by implementing model risk governance which equips them with means of identifying, quantifying, and mitigating the risks. 

The authors' careers span academic research, as well as model risk management and model development at financial institutions. Deriving from these exercises and building on similarities of model risk in finance and academic research, we believe the model risk management framework could be of immense benefit to academia and scholarship to inform practices in academic research and prevent reputational impacts. 

In this paper, we present three core principles of financial risk management and propose their application in academia. These principals are \emph{model ownership}, \emph{documentation}, and \emph{effective challenge}. We argue establishing a framework rooted in these principles in academia  would improve the quality of academic research and reduce errors in research projects. In some extreme cases, it would control for academic misconduct, and consequently, reduce substantial  financial costs incurred by academic institutions as a result.\footnote{For more on the financial cost and damages to the reputation of researcher as a result of fraudulent and erroneous research, see \cite{financialCost}.}

The life cycle of a financial model shares parallels with Thomas Kuhn's cycle of scientific revolutions; the latter is elaborated in his seminal work ``The Structure of Scientific Revolutions".  Just as Kuhn illustrated the progression of scientific paradigms, financial models undergo a similar evolution. In Section~\ref{s:cycle}, we provide a high-level description of the stages of the financial model life cycle, drawing parallels to Kuhn's framework to establish a foundational understanding for the subsequent discussion.

Moving forward, one of the core tenets of model risk management is accountability, which is manifested in the identification and assignment of model ownership. In Section~\ref{s:ownership}, we explore the concept of model ownership and the pivotal role of a model owner in financial institutions, proposing its equivalent in academia.

The main tool which enables model risk management is model documentation. The scope of model documentation spans the model lifecycle. In Section~\ref{s:documentation}, we discuss the use of documentation in financial institutions and then elaborate on the need for a similar use and scope of documentation in academia. 

Effective challenge serves as a cornerstone practice for identifying, quantifying, and mitigating model risk. Section~\ref{s:effective-challenge} delves into various aspects of this fundamental practice and suggests practical implementations for academia.

Finally, the paper concludes with Section~\ref{s:conclusion}, summarizing key insights and outlining potential avenues for future research.

\section{Model Lifecycle}\label{s:cycle}

The life cycle of a financial model draws inspiration from Thomas Kuhn's cycle of scientific revolutions. In his renowned book, "The Structure of Scientific Revolutions" (Kuhn, \cite{kuhn}), Kuhn outlined a basic progression cycle in scientific advancement, detailing the evolution of \emph{scientific paradigms} - a term he introduced in his book. The figure below provides a simplified depiction of this cycle.

\begin{center}
\includegraphics[scale=0.35]{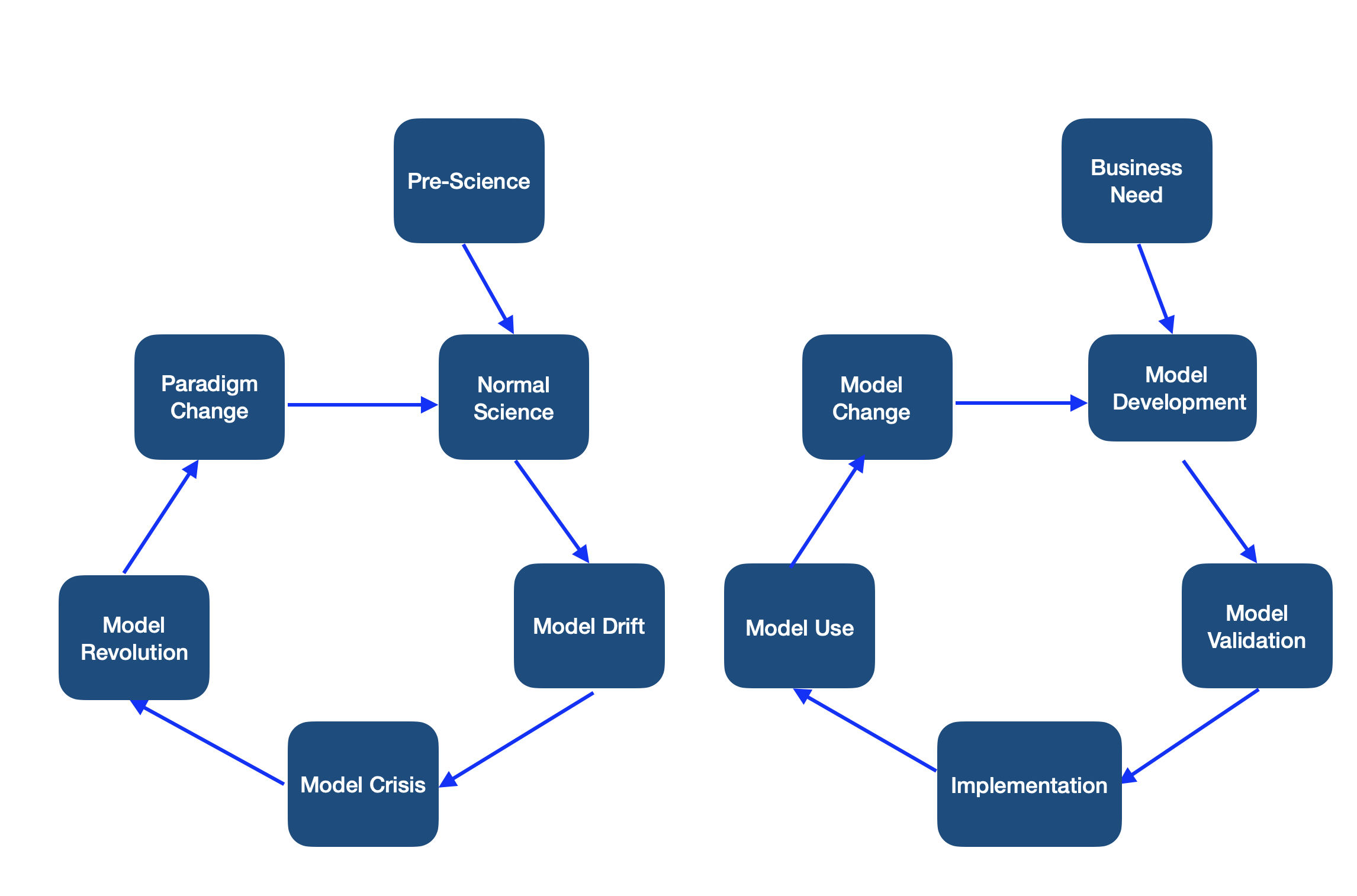}\\
Left: Kuhn's cycle of scientific revolution (Reference \cite{kuhn_cycle}),\\ Right: Financial Model lifecycle (Reference \cite{bennett})
\end{center}

The life cycle of a financial model follows a similar framework. The above figure also provides a visual representation of financial model life cycle. Beyond mere visual resemblance, those acquainted with the stages of Thomas Kuhn's cycle can readily discern similarities and strong analogies between the two cycles. In the rest of this section, we delve into the stages of the financial model life cycle to the extent necessary for comprehending the subsequent content.

The \emph{model owner}, is a unit or individual who is responsible for most of the model life cycle stages.  This encompasses activities such as selecting the model, coordinating its development, conducting initial testing, monitoring ongoing performance, analyzing outcomes, implementing changes, and crucially, maintaining comprehensive \emph{model documentation} (\cite{osfi}). Thorough documentation across all stages of the model's lifecycle is essential for effective model risk management; as stated earlier, it is the model owner's responsibility to develop and keep this documentation up to date. 

In the following, we briefly describe  the stages in the model lifecycle. The presented material is based on \cite{bennett}. 

\

{\bf Business Need:} Typically, the initial identification and articulation of the need for a financial model occur within the line of business. Following this, the line of business communicates the business requirement and encapsulates it in a business case, which guides the decision-making process regarding whether to commence development of a new model or to modify an existing one.

The business case should clearly define the model's purpose, specify the primary functional areas where the model will be utilized, and outline the necessary resources earmarked for the design and development phases.

\

{\bf Model Development:} Once the decision to commence model development has been made, the necessary personnel for this task will be identified. To ensure the resultant model aligns with requirements, the process should be guided by comprehensive consideration of all requirements, which include but are not limited to:

\begin{itemize} 
\item{{\bf Business requirements:} Encompassing the economic and financial context in which the model will operate, the business strategies it aims to implement, and the risk factors it aims to address. }
\item{{\bf Functional requirements:} Covering the specifications of the model's portfolio or scope, alongside any regulatory and legal mandates. } 
\item{ {\bf Technical requirements:} If specified at this stage, technical requirements may involve aspects such as precision level, processing speed, data prerequisites, and platform/interface functionalities. }
\end{itemize}

These requirements are closely linked to the data needs, which subsequently influence decisions regarding data sourcing. Once the data decisions are finalized, all relevant risks should be thoroughly documented and analyzed.

\

{\bf Model Validation (Effective Challenge):} \emph{Effective challenge} is ``critical analysis by objective, qualified, independent parties'' (\cite{feds}). While effective challenge principle is different risk management environments such as audit, in the modelling space, the focus is on scrutinizing the model development process and assessing the resulting model's appropriateness for its intended use. To allocate their model validation resources more efficiently, financial institutions first assess the risk of their existing models to determine the necessary level of rigour and the frequency of the effective challenge. The effective challenge should ensure that the models meet their intended purpose and adhere to applicable usage restrictions.

Independence stands as a fundamental principle within the framework of an effective challenge. Validation teams must remain uninvolved in the ownership, development, or utilization of a model to ensure their ability to operate independently and avoid conflicts of interest. Before implementation, all models falling under the purview of a financial institution's model risk management must undergo validation and approval.  

{\bf Model Utilization (Implementation/Model Use/Model Change):}  Upon approval, the a model will be implemented for model use. The \emph{implementation} process shall include thorough testing and documentation. Any testing and analysis results confirming the satisfactory implementation must be documented and subject to effective challenge. Once the implementation process is approved, the model may be utilized by the line of business for its originally intended purposes.

During  \emph{model use}, continuous monitoring and documentation of the model output and performance are required. If the model performance does not meet the acceptability requirements that are established during model validation, model use must cease, and the model should undergo review and if necessary improvement processes. This could, among other factors, necessitate a \emph{model change}.

A business case for a model change would initiate the development of a new model. This process entails model development, validation, and implementation before it can be utilized by the line of business.

\section{Ownership}\label{s:ownership}

As stated above, model ownership is one of the pillars of model risk management in financial institutions. It is implemented through accountability over model risk. In this section, we introduce a model owner role in the financial system (Subsection~\ref{ss:owner-fi}). Subsequently, we discuss the benefits of defining and establishing the role of ownership for academic research (Subsection~\ref{ss:owner-ar}).

\subsection{Model Ownership in Financial Institutions:}\label{ss:owner-fi}

The majority of modes within a financial institution are  in support of business decisions, but there are others, such as the ones used for marketing. Ownership of the former is defined and assigned to an officer who is a member of the management team. This owner is responsible for compliance of the models with the financial institution’s model risk governance. He or she assumes the responsibility for model risk associated with the model at every stage of its lifecycle. To meet this requirement, the model owner ensures the artifacts and activities that pertain to the model, such as model data, methodology, testing and monitoring, are conceptually sound, appropriate for the model purpose, and comprehensively documented (see Section \ref{s:documentation}). The latter enables transparency and replicability of model development, testing and, monitoring through persistence and availability of artifacts corresponding to each of these stages in the model lifecycle.

The model owner also liaises with the other model stakeholders such as the development team, validation, and internal audit, etc. To be able to execute all these requirements, the ownership must have no conflict of interests; in particular, the owner cannot be a member of the validation or audit teams who provide effective challenge (see Section \ref{s:effective-challenge}). 

\subsection{Ownership of Academic Research: }\label{ss:owner-ar}
The concept of research ownership has not been explicitly defined. The closest analogue of an owner of an academic research project is the corresponding author whose assumed responsibility is to undertake any communications with the journal where the paper has been submitted. Building on the concept of a model owner in financial institutions, one could propose a range of responsibilities to be assumed by the owner of academic research beyond the publication correspondence. 

To properly translate the concept of model ownership to academic research, we propose to assign such ownership to the senior scholars who oversees the research project, supervises the step by step execution (i.e. throughout the research lifecycle), and remains the main point of contact regarding the methodology, input and output data, and eventual publication of the results.  Again, following the paradigm of a financial institution, we propose that the owner of the academic research project be expected to assume responsibility over the data, methodology, testing and monitoring, and all related decisions.

With an ownership role explicitly defined and assigned for a research project, and given the above mentioned responsibilities regarding data decisions and methodology assumptions, the owner would  ensure the data and applied methodologies are sound, appropriate for the purpose, and replicable. Consequently, one could expect that many potential fatal flaws in research projects would have been prevented. And in the unlikely but possible case of erroneous results, the owner would have facilitated the remedial actions. 

There are several incidents and allegations in academic research which would highlight the immediate benefits of  centralization of such responsibilities, specially in cases where  erroneous results were alleged to have been  published. Here we could mention  a highly cited paper which was retracted (\cite{gino}) following the validity of the research data  being questioned. 
According to a news coverage of the incident (\cite{ArielyGineData}), four of the five authors indicated that they played no part in collecting the data. Arguably, such ambiguity in the identifying party responsible for the data would hinder a thorough investigation in the validity of the results. A research owner, as described above, would have the knowledge of the responsible parties for each aspect of the project. Therefore, after discussing the allegations with the responsible team members and conducting necessary internal investigations, the owner would have been able to clarify whether the  concern raised with the data is valid. And in the cases of the allegations  not being accurate or  not materially impacting the results, the owner would provide evidence.

\section{Documentation} \label{s:documentation}

It would be impossible to imagine the implementation of model risk management framework without detailed model documentation.  Model documentation is crucial at every stage of the model lifecycle. In this section, we first discuss the structure and function of model documentation for financial entities (Subsection~\ref{ss:documentation-fi}). We then highlight the benefits of a similar documentation practice for academic research (Subsection~\ref{ss:documentaiton-ar}). 

\subsection{Model Documentation in Financial Institutions:} \label{ss:documentation-fi}
Documenting a model is one of the fundamental tasks of a model developer. Comprehensive model documentation spans not only development stages such as data collection, methodology selection, algorithm development/coding, and testing but also post-development stages such as model implementation, its use in production, i.e. to support business decision, as well as validation. Thus, documentation is a foundational component model risk management in financial institutions. 

In the following we discuss four components in further details which have their counterparts in Subsection~\ref{ss:documentaiton-ar}. 
\begin{enumerate}
\item{{\bf Model Data:} Detailed description of  modelling data (i.e., any data used for development and testing of the model) is provided in the model documentation.  In particular, the documentation discusses appropriateness of the available and selected data, any data quality issues, as well the use of their mitigates such as data proxies, if applicable. Likewise, missing data, outliers, and biased data and their mitigating treatments are covered. Furthermore, model developers justify data selection and treatment decisions, such as data proxies representativeness. }
\item{{\bf Modelling Methodology:} The methodology description is the centrepiece of model documentation. It elaborates on the suitability of the proposed methodology compared to alternatives. It articulates the analysis conducted in support the methodology selection including qualitative and quantitative arguments as well as the literature review. 

The documentation must be detailed enough to support replication of the development steps by a third party. Hence, if applicable, the programming code needs to be thoughtfully documented.  }
\item{{\bf Model Testing:} A financial institution with mature model risk practice would stipulate the minimum testing required along with acceptability criteria within the internal model risk management policies and guidelines. Testing ensures the satisfactory performance of a model which is developed and used within the enterprise. In model documentation, model developers report on adherence to these testing requirements as well as any extra tests and analyses which are deemed necessary to ensure soundness of the underlying model. Test outcomes and their interpretation are captured in the documentation.  }
\item{{\bf Model Limitations and Outputs:} The documentation defines model outputs and their  use. To do so, the documentation includes a description of the model outputs and their properties such as data definitions, acceptability ranges, and limitations. {\emph{need more on limitations}}}
\end{enumerate}

\subsection{Documentation in Academic Research}\label{ss:documentaiton-ar}
A research project in academia often leads to peer reviewed journal publications. This publication aims to summarize the research project and state the main results. With an exception of some abstract subjects, a journal publication does not typically provide a complete account of the work conducted, the inputs, methodology, and limitations, nor is it intended to serve as complete documentation for the project. 

Compared to comprehensive documentation for models in financial institutions, journal papers leave out the details of many aspects of a research project. Given all the benefits of model documentation, as discussed in Subsection~\ref{ss:documentation-fi}, we recommend that the researchers engaged in a research projects be mandated to document their work to a more rigorous standard. Such documentation would fill the gaps present in journal papers. Peer review prior to publishing  enabled by comprehensive and rigorous documentation would serve to prevent publication of erroneous and fraudulent results. 

Modelled on the scope of model documentation in a financial institution as described above, complete academic research project documentation must  cover research data and methodology as well as their respective limitations, model testing and outcomes. These categories are analogous to the ones discussed in Subsection~\ref{ss:documentation-fi}, namely, model data, model methodology, model testing, and model outputs, respectively. In the following, we discuss these categories in the context of academic research while leveraging the arguments  of Subsection~\ref{ss:documentation-fi}.

\begin{enumerate}
\item{{\bf Research Data:}  A quick review of the list of retracted papers on \href{https://retractionwatch.com}{Retraction Watch website} would reveal that most of the retractions are due to either allegedly fabricated or fraudulent data, or non-replicability of research data. This kerfuffle could be remedied if the researchers were required  to  provide a comprehensive research document which covers all aspects of the data used in their projects, including the source of data, data dictionary, data treatments and modifications, such the use of data proxies, as well as data quality issues and other limitations. Furthermore, the documentation must clearly justify the representativeness of the used data for the purposes of the project. 

We believe the researchers' attempts to populate this part of a research document, as described above, would lead to a notable decrease in the chance of data issues, such as ``data dredging".   }

\item{{\bf Research Methodology:}  Here, we use the word ``methodology" to address two aspects of inductive research. Recall that inductive learning is archived through generalization of experience, i.e. data.

\begin{itemize}
\item{{\bf Decisions made on the selection of statistical and mathematical methodologies} which are used in inductive research must be documented. This applies to a range of methodologies from a simple hypothesis testing (relying on $p$-value) to a novel unsupervised machine learning algorithm. 

A clear explanation of the rationale of such methodologies must be stated, and their appropriateness discussed. One could argue that documenting the methodology would overcome many of the shortcomings observed in scientific research projects which are due to lack of diligence in choosing the most appropriate statistical and mathematical tests and algorithms in conducting the project.  }
\item{{\bf Algorithmic or automated realization of the methodology} (e.g., in a spreadsheet or in programming code) would also require thorough documentation. This section of the documentation is vital, in particular, for replicating research outcomes. It is no secret that many retracted journal papers are pulled  on the ground of the results not being replicable, whether by the authors or by other researchers in the field.\footnote{As an example for this case, look at the retraction note for \cite{roos}.} Proper documentation of the methodology realization would prevent these cases. 

Since the replicability is an important tool in detecting erroneous and fraudulent research (see Section~\ref{s:effective-challenge}), this aspect of the methodology documentation  is a foundational instrument in research projects passing effective challenge.} 
\end{itemize}
}
\item{{\bf Hypothesis Testing and Limitations:} Since hypothesis testing is one of main statistical tools in most of inductive research projects, it is necessary to clearly document any hypothesis testing conducted in the course of a research project. In particular, the researchers must clearly state their hypotheses, testing methodology, and its limitations. In addition, articulating any limitations to their testing and methodology are required to be thoroughly  documented. 

As discussed in \cite{p-hacking}, the presence of  false positive results in inductive studies could account for a notable portion of published research. For example, reliance on $p$-values only is one of the sources of false positives. In general, the researchers are expected to conduct sufficient testing to ensure such limitations are addressed, and document any remaining limitations following from the selected methodology. }
\item{{\bf Results:} 
Finally, the documentation shall articulate the correct interpretation of the research results. This discussion is based on all components of the project, such as input data, methodology, testing, and etc., and their limitations. In this part of the documentation, the researchers would state all the caveats regarding the use and understanding of the outcomes. 
}
\end{enumerate}

\section{Effective Challenge}\label{s:effective-challenge}

Effective challenge is  the  method for identifying and mitigating model risk. Drawing from the longstanding practices of financial institutions (see Section 6 in \cite{bennett}), we propose a similar approach for academic research to detect fallacies, misconduct, fraudulent activities, plagiarism, and unsupported findings. Implementing this practice not only helps academia reduce the occurrence of suboptimal outcomes but also mitigates reputational and financial risks for academic establishments and individual scientists (see \cite{financialCost}). Moreover, the impact of erroneous or fraudulent research extends beyond financial and reputational damage, as it can have real-world implications for society and the environment. For example, a prominent teaching center affiliated with Harvard Medical School is currently pursuing the retraction of six studies and corrections to an additional 31 papers. These actions come in response to allegations that certain senior researchers falsified data (see \cite{cancer}). While we have not followed up whether there has been any practical implementation of this medical research in the publications mentioned above, one can see that such research misconduct could have devastating effects on the lives of cancer patients.

Similarly to financial institutions, we first propose a rating mechanism to prioritize projects with the highest impact, thereby increasing the effectiveness of this practice. Subsection~\ref{ss:rating} provides a summary of the risk rating methodology in financial institutions and proposes an analogous approach for academic research.

Next, in Subsection~\ref{ss:validation}, we discuss the validation function, beginning with its role in financial institutions and then examining its  relevance for academia.

\subsection{Risk Rating}\label{ss:rating}

\noindent{\bf Risk Rating Models:} The risk management frameworks of  financial institutions are developed to prioritize risks, primarily through the concept of \emph{model risk rating}.  A model's risk rating is determined by assessing the potential financial and reputational impact of a model error, considering two key dimensions: the materiality of the portfolio subject to modelling and the magnitude of possible errors inherent in the model. Models rated high in this risk rating scheme are given priority attention and would undergo higher level of scrutiny to identify and mitigate their model risk and its impact on the financial entity.  While we will explore  model validation in greater detail in Subsection~\ref{ss:validation}, we propose a similar rating scheme for academic research.

\noindent{\bf Rating Research Projects:} A rating system for assessing the risk level of academic research should consider various factors, including financial, reputational, environmental, and social impacts. In other words, academic research with greater environmental or social implications should be prioritized for validation and subjected to a higher level of scrutiny.

We believe that hosting institutions, similarly to financial ones, should develop rating guidelines to evaluate research projects. Academic institutions should collect descriptive information on their projects to facilitate this rating process. This information would be included in an inventory of the projects and their attributes.

Establishing such research databases would enhance visibility and monitoring of research activities, particularly those requiring review by the university's institutional ethics boards. For further motivation, consider the case of the CRISPR gene-edited babies \cite{CRISPR}, which was conducted without proper ethical review and approval. The project was universally condemned by experts and resulted in criminal charges for some of the researchers  involved researchers. If this project and its metadata had been tracked through a research database and validated based on its attributes, red flags would have been raised  early and potentially preventing the damage.

\subsection{Validation}\label{ss:validation}

Validation is the cornerstone of model risk management in financial institutions, ensuring that models are sound, meet requirements, and serve their intended purposes. The significance of the model validation function became even more apparent following the credit crisis of 2008-2009. In particular, Salmon \cite{2008Models} demonstrated how reliance on the Gaussian copula assumption in financial models, which estimated the probability of losses in a pool of loans or bonds, contributed to the global financial crisis. This crisis unfolded despite the awareness of the limitations of the modelling assumption, as warnings were largely ignored by the financial system. With a properly operationalized model validation function and model validation displine enforced, the performance on the models built on the Gaussian copula assumption would be  closely monitored, thereby allowing the mitigation of model risk.

In this section, expanding upon the high-level description of the model validation/effective challenge process outlined in Section~\ref{s:cycle}, we delve into its implementation within financial institutions. Subsequently, we investigate how these practices could be adopted by and applied to academia, with the aim of enhancing the quality of academic research.

\;

\noindent {\bf Validation of  Models in Financial Institutions:}
To ensure that model validation is not adversely impacted by conflicts of interest, it is independent from  model development  and ownership. The validation team in a financial institution often consists of qualified personnel who have no direct or indirect interests in the development and implementation of any models.

The validators are responsible for reviewing and approving newly developed models or models undergoing the model change process (see Section~\ref{s:cycle}). To accomplish this, the validation team conducts thorough investigations into the soundness and appropriateness of the models. These activities include, but are not limited to:
\begin{itemize}
\item {\bf Model Replication:} The validation team replicates the logic, operations, and testing of the model. They rely on the detailed description of the development and testing  provided in the model documentation (see Subsection~\ref{ss:documentation-fi}) to ensure accurate replication aiming to reproduce the development results.
\item {\bf Theoretical Review:} The validation team reviews the modeling methodology to ensure its conceptual and theoretical appropriateness. Validators carefully assess the methodology, understanding its underlying limitations and shortcomings. They rationalize their assessment and complete the list of limitations, if needed. 
\item {\bf Data Validation:} This aspect of validation may involve various exercises to assess the quality and representativeness  of modeling and testing data. Validators thoroughly investigate any data treatment, manipulation, and limitations.
\item {\bf Model Testing:} In addition to reviewing and replicating the test results of model developers documented in the model documentation, validators may conduct further testing and benchmarking exercises. This provides further assurance that the model and its outputs meet the business requirements, with documented claims regarding model perfromance being factual.
\end{itemize}

The validation process results in a validation report that discusses all findings. In cases where material flaws are identified, the model is returned for redevelopment. A rejected model is not used until the issues raised by the validation team are resolved. To address this, the model developers update the model accordingly, and the revised version is submitted for another round of validation. Subsequently, the revised validation report should reveal no material findings.

\;

\noindent {\bf Validation of Academic Research Results:} While the referee process of scientific journals intends to address validations of research coming for publication, in practice, it often falls short of the minimum standards required for validation in financial institutions. In particular, with the rise of AI-generated content, the shortcomings of the current peer review process have been highlighted, particularly in its failure to flag pseudo-scientific material generated by AI. For example, see \cite{mouse}, where a peer-reviewed science journal published a paper containing nonsensical AI-generated images, including distorted text and an inaccurately depicted diagram of a rat anatomy. 

Drawing on the successful practices of model validation in financial institutions, we propose that a similar function  be developed for academic research. This function would include components such as replication, methodological review, data validation, and additional statistical analyses, which are analogous to model replication, theoretical review, data validation, and model testing discussed earlier for model validation in financial institutions.

\begin{itemize}
\item{{\bf Replication:}  
Many scientific papers are retracted due to  inability to reproduce their results. A thorough and independent replication of a research project, conducted by experts in the field, should ensure the integrity of the data and its processing, including any requirements, limitations, and treatments. Such validators can also identify errors in the execution of the work by reconstructing the code and methodology. }

\item{{\bf Methodology and Test Selections:} To conduct their research and draw conclusions, researchers select methodologies and tests deemed appropriate for their work. These research decisions would benefit from independent validation. Validators, drawing on their knowledge and expertise, would verify whether researchers have made appropriate decisions and accounted for the limitations of selected methodologies and tests.}

\item{{\bf Data Validation:} We discussed the importance of data in Section~\ref{s:documentation}. A validation  would include a comprehensive investigation of the research data. This involves the following tasks:\begin{itemize}
\item{Vetting the research data and assessing its relevance for the intended purpose.}
\item{Verifying data definitions and mappings.}
\item{Assessing any data gaps, outliers, and  treatments to mitigate them.}
\item{Verifying the rationale and appropriateness of any data proxies.}
\item{Assessing the sample size and its credibility  and studying any biases and data scarcity.}
\end{itemize}
}

\item{{\bf Further  Statistical Analyses :} In addition to reviewing and replicating the statistical testing results, validators of academic research may need to conduct extra analyses and benchmarking to ensure the defendability of the conclusions and identify any limitations introduced by the researchers. This may involve using alternative datasets and methodologies for verification. }
\end{itemize}

It is worth highlighting that the aforementioned assignments and tasks of research validation would not be feasible without comprehensive documentation, as detailed in Subsection~\ref{ss:documentaiton-ar}. Furthermore, having an assigned ownership role for a research project, as discussed in Subsection~\ref{ss:owner-ar}, would significantly facilitate the process and improve its transparency and efficiency.

\section{Proposed Solution for Academia}\label{s:resources}

The majority of regulated financial institutions are required to develop and advance policies and guidelines for managing model risk, which are operationalized  through the function  dedicated to managing model risk.\footnote{For example, in Canada, the Office of the Superintendent of Financial Institutions has issued Guideline E23, 'Enterprise-Wide Model Risk Management for Deposit-Taking Institutions,' outlining expectations for institutions to establish sound policies and practices for an enterprise-wide model risk management framework.}
The team supporting this function typically consists of validation and governance branches. We briefly discussed some responsibilities assigned to model validation  in Section~\ref{ss:validation}. The governance team, at a high level, oversees all logistics required for successful  implementation of model risk management policies and guidelines, such as maintaining a model inventory and rating models for their risk (see Subsection~\ref{ss:rating}). 

The size of a model risk management team corresponds to the size of financial institutions and the extent of their reliance on financial models.  While model risk management teams are an established and well-resourced part of the enterprise risk management infrastructure in major financial institutions, academic institutions often lack the resources necessary to establish such infrastructures and conduct independent effective challenge.

Based on our observations during our academic careers and ongoing outreach to academia for our daily work assignments, in this section we propose a potential resource model that could support validation of  research conducted in academia. Specifically, we suggest the development of governance and independent review within each academic institution that mirrors aspects of model risk management seen in financial institutions.

\begin{itemize}
\item{{\bf Governance:}  Academic institutions would establish a group that functions similarly to the governance structure of model risk management teams. This group would maintain a comprehensive inventory of all active research projects ongoing within the institution and rank them based on principles  outlined in Subsection~\ref{ss:rating}. Based on the intake and ranking, this group would request resources for independent review to projects, taking into account  independence, knowledge, and expertise of review teams.}

\item{{\bf Independent Review:} 
We propose that academic institutions designate review teams to conduct independent review of each research project, independent from the researchers who developed them. The scope of the review would be determined by the aforementioned risk rating.

Acknowledging the limited resources available to academic institutions, we suggest that review teams be comprised of graduate students, particularly Ph.D. candidates. These students would work under the supervision of a faculty member, who would act as the team leader. 
Apart from the benefit of having the research project reviewed and validated, this process would provide  students participating  in each review  with valuable learning opportunities, exposing them to the latest ongoing research in their field. Moreover, to conduct thorough investigations, graduate students would need to perform literature review and understand the limitations of various methodologies and testing analyses. Engaging in data validation would provide an excellent opportunity for them to learn about research datasets, data limitations, and treatments conducted to mitigate data gaps and outliers. Additionally, students would need to perform further analyses and testing to ensure the accuracy and relevance of the claimed results in the underlying research, which would require them to familiarize themselves with a variety of testing and analysis methods.

Beyond acquiring skills and knowledge through these tasks, this assignment would cultivate a culture of prudence and trustworthy academic research in the future generation of scientists who will lead the frontiers of science for years to come.}

\end{itemize}

\section{Conclusion}\label{s:conclusion}
In this paper, we proposed and outlined a model  designed to detect and prevent erroneous and fraudulent academic results. The proposal is informed by model risk management practices used by financial institutions to identify and mitigate model risk.

While we believe our proposals would help academic institutions improve the quality and accuracy of published results, we recognize that implementing these principles and practices would require further study and more granular design decisions. While seemingly increase the burden on academic institutions, the resulting discipline would protect scientists and scientific establishments from potential adverse outcomes of human error. 

\;

\noindent{\bf Authors' note:} The views expressed in this paper are the authors’ own and do not necessarily reflect the views of the Office of the Superintendent of Financial Institutions (OSFI).

The authors did not conduct investigations into the exemplified research publications and projects mentioned in this manuscript; these retracted publications were solely mentioned based on and to the extent covered by the cited references listed below. The accuracy of these references is not known to the authors, and they do not make any claims regarding these statements and allegations; therefore, the authors do not bear any responsibility and shall not be held liable for them.

Proposed methodologies and solutions in this manuscript are solely based on the success of the existing corresponding practices in financial institutions.

Finally, the authors would like to thank Ali Yazbek for drawing their attention to \cite{plnetMoney} in the first place.


\end{document}